\documentclass[nohyper,notoc]{article} 
\usepackage{epsfig}
\input epsf
\def\beq{\begin{equation}}
\def\eeq{\end{equation}}
\def\be{\begin{equation}}
\def\ee{\end{equation}}
\def\bea{\begin{eqnarray}}
\def\eea{\end{eqnarray}}
\def \lsim{\mathrel{\vcenter
     {\hbox{$<$}\nointerlineskip\hbox{$\sim$}}}}
\def \gsim{\mathrel{\vcenter
     {\hbox{$>$}\nointerlineskip\hbox{$\sim$}}}}
\def\gappeq{\mathrel{\rlap {\raise.5ex\hbox{$>$}}
{\lower.5ex\hbox{$\sim$}}}}
\def\lappeq{\mathrel{\rlap{\raise.5ex\hbox{$<$}}
{\lower.5ex\hbox{$\sim$}}}}

\def\Huv{\langle H_u \rangle}

\begin{document}
\vspace*{-1in}
\renewcommand{\thefootnote}{\fnsymbol{footnote}}
\begin{flushright}
OUTP-02-10P \\
IPPP/02/16  \\
DCPT/02/32
\end{flushright}
\vskip 5pt
\begin{center}
{\Large {\bf A lower bound on the right-handed neutrino
mass \\ from leptogenesis
}}
\vskip 25pt
{\bf Sacha Davidson$^*$ and
Alejandro Ibarra$^{\dagger}$}
 
\vskip 10pt 
 {\it $^*$ Dept of Physics, University of Durham,
Durham, DH1 3LE, United Kingdom}\\
{\it $^{\dagger}$ Theoretical Physics, University of Oxford,
 1 Keble Road, Oxford, OX1 3NP, United Kingdom}\\
\vskip 20pt
{\bf Abstract}
\end{center}
In the seesaw model, the baryon asymmetry of the Universe can be
generated by the decay of the lightest right-handed neutrino,
$\nu_{R}$. For a hierarchical spectrum of right-handed neutrinos, we
show that there is a model independent upper bound on the CP asymmetry
produced in these decays: $\epsilon< 3~m_{\nu_{3}} M_{\nu_{R}}/ (8 \pi
\Huv^2)$.  This implies that $\epsilon$ and the mass $ M_{\nu_{R}}$
of the lightest right-handed neutrino are not independent parameters,
as is commonly assumed.  If $m_{\nu_{3}} = \sqrt{\Delta m^2_{atm}}$
and the $\nu_R$ are produced thermally, then leptogenesis requires a
reheat temperature of the Universe $T_{reh} > M_{\nu_R} > 10^8 $
GeV. Reasonable estimates of $\nu_R$ production and the subsequent
washout of the asymmetry, as made by Buchm\"uller and Pl\"umacher,
imply $M_{\nu_R} > 10^9 $ GeV, and $T_{reh} > 10^{10}$ GeV.
Implications for the gravitino problem are also discussed.
\\

PACS number(s):~12.60.Jv, 14.60.Pq, 98.80.Bp\\

\begin{quotation}
  {\noindent\small 

\vskip 10pt
\noindent
}

\end{quotation}

\vskip 20pt  

\setcounter{footnote}{0}
\renewcommand{\thefootnote}{\arabic{footnote}}
\section{Introduction}

The discovery of neutrino oscillations has been one of the most
exciting experimental results in the last years. Although a deficit in
the flux of solar neutrinos, observed for the first time in the late
60s \cite{Cleveland:1998nv}, suggested that neutrinos oscillate, it
was not until the Super-Kamiokande experiment \cite{SK} that the
oscillation hypothesis acquired strength\footnote{Other hints for
neutrino oscillations have been reported in \cite{experiments}.}.  As
is widely known, these results are nicely explained if neutrinos have
a small mass. Explaining why the neutrino mass scale is so small is
one of the current unsolved problems in particle physics.  An elegant
solution is the see-saw mechanism \cite{seesaw}, which consists on
adding a heavy Majorana fermion per generation to the Standard Model
(SM) particle content, singlet with respect to the SM gauge group, and
coupled to the Higgs doublet through a Yukawa coupling.  Although very
appealing, this model suffers from a serious hierarchy problem, since
the right-handed neutrinos produce a (large) quadratically-divergent
radiative correction to the Higgs mass. Hence, in this letter we will
concentrate on the supersymmetric version of the see-saw mechanism,
that stabilizes the Higgs mass against the dangerous quadratic
divergences that otherwise appear.

The leptonic part of the corresponding superpotential reads
\bea
\label{superp}
W_{lep}= {e_R^c}^T {\bf Y_e} L\cdot H_d + {\nu_R^c}^T {\bf Y_\nu}
L\cdot H_u - \frac{1}{2}{\nu_R^c}^T{\cal M}\nu_R^c,
\eea
where $L_i$ and $e_{Ri}$ ($i=e, \mu, \tau$) are the left-handed lepton
doublet and the right-handed charged-lepton singlet, respectively, and
$H_d$ ($H_u$) is the hypercharge $-1/2$ ($+1/2$) Higgs doublet.  ${\bf
Y_e}$ and ${\bf Y_{\nu}}$ are the Yukawa couplings that give masses to
the charged leptons and generate the neutrino Dirac mass, and $\cal M$
is a $3 \times 3$ Majorana mass matrix that does not break the SM
gauge symmetry.

It is natural to assume that the overall scale of $\cal M$, denoted by
$M$, is much larger than the electroweak scale or any soft mass.
Therefore, at low energies the right-handed neutrinos are decoupled
and the corresponding effective Lagrangian contains a Majorana mass
term for the left-handed neutrinos:
\bea  \delta {\cal L}_{lep}={e_R^c}^T {\bf Y_e} L\cdot H_d 
-\frac{1}{2}\nu^T{\cal M}_\nu \nu + {\rm h.c.},
\eea
with
\bea
\label{seesaw}
{\cal M}_\nu= {\bf m_D}^T {\cal M}^{-1} {\bf m_D} = {\bf Y_\nu}^T
{\cal M}^{-1} {\bf Y_\nu} \langle H_u^0\rangle^2, \eea
giving neutrino masses suppressed with respect to the typical fermion
masses by the inverse power of the large scale $M$.

Another attractive feature of the see-saw mechanism is that it
provides a natural mechanism \cite{Fukugita:1986hr} 
to generate the observed Baryon Asymmetry
of the Universe (BAU) \cite{Buchmuller:2000wq}, which we parametrize
as $\eta_B = (n_B - n_{\bar B})/s$, where $s$ is the entropy
density. This quantity, that governs the abundances of the light
nuclei D, $^3$He, $^4$He and $^7$Li, must have a value in the range
$\eta_B \simeq (0.3-0.9) \times 10^{-10}$ to reproduce the observed
abundances \cite{Olive:2000ij}, according to the Standard Big Bang
Nucleosynthesis scenario.  The generation of this baryon asymmetry
requires three ingredients: baryon number violation, C and CP
violation, and a deviation from thermal equilibrium
\cite{Sakharov:1967dj}. All these conditions are fulfilled in the out
of equilibrium decay of right-handed neutrinos \cite{Fukugita:1986hr}
and sneutrinos in the early Universe. For conciseness, and since we
are concerned only with supersymmetric leptogenesis, in what follows
we will use right-handed neutrinos, and the shorthand notation
$\nu_R$, to refer both to right-handed neutrinos and right-handed
sneutrinos.

Let us briefly review the mechanism of generation of the BAU through
leptogenesis \cite{Fukugita:1986hr,Buchmuller:1999cu,Hirsch:2001dg}.
At the end of inflation, a certain number density of right-handed
neutrinos, $n_{\nu_R}$, is produced, that depends on the cosmological
scenario. These right-handed neutrinos decay, with a decay rate that
reads, at tree level,
\beq
\Gamma_{D_i}= \Gamma(\nu_{R_i} \rightarrow  \ell_i H_u) + \Gamma
({\nu}_{R_i} \rightarrow  \tilde{L}_i \tilde{h}_u)
= \frac{1}{8 \pi} ({\bf Y_{\nu} Y_{\nu}}^{\dagger})_{ii} M_i.
\label{decay-rate}
\eeq
The out of equilibrium decay of the lightest right-handed neutrino
$\nu_{R_1}$ creates a lepton asymmetry given by
\beq
\eta_{L} = \frac{n_\ell - n_{\bar \ell}}{s} = 
 \frac{n_{\nu_R} +n_{\tilde{\nu}_R} }{s} ~ \epsilon ~ \delta.
\label{etaL}
\eeq
The value of $(n_{\nu_R} +n_{\tilde{\nu}_R})/s$ depends on the
particular mechanism to generate the right-handed (s)neutrinos. On the
other hand, the CP-violating parameter
\beq
\epsilon_i = \frac{\Gamma_{D_i}- \bar \Gamma_{D_i}}
{\Gamma_{D_i}+ \bar \Gamma_{D_i}},
\label{epsilon}
\eeq
where $\bar \Gamma_{D_i}$ is the CP conjugated version of
$\Gamma_{D_i}$, is determined by the particle physics model that gives
the masses and couplings of the $\nu_R$.  Finally, $\delta$ is the
fraction of the produced asymmetry that survives washout by lepton
number violating interactions after $\nu_R$ decay.  To ensure $\delta
\sim 1$, lepton number violating interactions (decays, inverse decays
and scatterings) must be out of equilibrium when the right-handed
neutrinos decay. This corresponds approximately to $\Gamma_{D_1}<H|_{T
\simeq M_1}$, where $H$ is the Hubble parameter at the temperature
$T$, and can be expressed in terms of an effective light neutrino mass
\cite{Buchmuller:1999cu,Plumacher:1997kc}, 
$\widetilde m_1$,  as
\beq
\widetilde{m}_1 = 
\frac{ 8 \pi \langle H_u^0\rangle ^2}{M_1^2}
\Gamma_{D_1} = 
({\bf Y_{\nu} Y_{\nu}}^{\dagger})_{11} 
\frac{\langle H_u^0\rangle ^2}{M_1}
\lappeq 5 \times 10^{-3} {\rm eV}.
\label{mtilde}
\eeq
This requirement has been carefully studied
\cite{Buchmuller:1999cu,Plumacher:1997kc,Barbieri:2000ma}; the precise
numerical bound on $\widetilde m_1$ depends on $M_1$, and can be found
in \cite{Plumacher:1997kc}.

The last step is the transformation of the lepton asymmetry into a
baryon asymmetry by non-perturbative B+L violating (sphaleron)
processes
\cite{Kuzmin:1985mm}, giving 
\beq
\eta_{B} =\frac{C}{C-1} \eta_{L},
\label{BAU}
\eeq
where $C$ is a number ${\cal O}(1)$, that in the Minimal
Supersymmetric Standard Model takes the value $C=8/23$.

Although the supersymmetric leptogenesis scenario is a very attractive
framework to generate the BAU, it is not free of problems. Namely,
there is a potential conflict \cite{Buchmuller:1999zf} between the
gravitino bound \cite{gravitino,ENS} on the reheat temperature, and the
thermal creation of right-handed neutrinos. In a plasma at high
temperature, gravitinos are abundantly produced, and their late decay
could modify the abundances of light nuclei, contrary to observation.
This sets an upper bound on the reheat temperature that will have an
important role in our discussion.

\section{Upper bound on the CP asymmetry}
\label{sec2}

As was explained in the introduction, the out of equilibrium decay of
$\nu_R$s generates a lepton asymmetry that is proportional to the CP
asymmetry, $\epsilon$. To compute the CP asymmetry, it is convenient
to work in the flavour basis in which the charged-lepton Yukawa
matrix, $\bf{Y_e}$, and the gauge interactions are flavour-diagonal
(therefore, all the lepton flavour mixing is in $\bf{Y_{\nu}}$).  In
this basis, the neutrino mass matrix can be diagonalized by the MNS
\cite{Maki:1962mu} matrix $U$ according to
\be
\label{Udiag}
U^T{\cal M}_\nu U={\rm{diag}} (m_1,m_2,m_3)\equiv D_m, 
\ee
where $U$ is a unitary matrix that relates flavour to mass eigenstates
\bea  \pmatrix{\nu_e \cr \nu_\mu\cr \nu_\tau\cr}= U \pmatrix{\nu_1\cr
\nu_2\cr \nu_3\cr}\,.
\label{CKM}
\eea
On the other hand, one can always choose to work in a basis of
right-handed neutrinos where ${\cal M}$ is diagonal
\be {\cal M}={\rm{diag}} (M_1,M_2,M_3)\equiv
D_M, 
\ee
with $M_i\geq 0$. In this basis, the CP asymmetry can be readily
computed, yielding the result
\bea
\epsilon_i  \simeq  - \frac{1}{8 \pi} \frac{1}{[{\bf Y_{\nu} Y_{\nu}}
^{\dagger}]_{ii}}  \sum_j {\mathrm{Im}}
 \left\{ [{\bf Y_{\nu} Y_{\nu}}^{\dagger}]^2_{ij} \right\}
f \left(  \frac{M_j^2}{M_i^2} \right),
\eea
where \cite{Covi:1996wh}
\beq
f(x) = \sqrt{x}  \left( \frac{2}{x - 1 } + \ln \left[ \frac{1+x}{x} \right]
\right).
\eeq
Here, we will assume that the masses of the right-handed neutrinos are
hierarchical\footnote{See \cite{Pilaftsis:1998pd} for a discussion on
leptogenesis with degenerate right-handed neutrinos.}. In this case,
the lepton asymmetry is essentially generated in the decay of the
lightest right-handed neutrino, so is proportional to
\bea
\epsilon_1 \simeq  -\frac{3}{8 \pi} \frac{1}{[{\bf Y_{\nu} Y_{\nu}}
^{\dagger}]_{11}} \sum_j {\mathrm{Im}} \left\{ [{\bf Y_{\nu}
Y_{\nu}^{\dagger}}]^2_{1j} \right\} \left( \frac{M_1}{M_j} \right)
= - \frac{3}{8 \pi} \frac{M_1}{\langle H_u^0\rangle ^2}\frac{1}
{[{\bf Y_{\nu} Y_{\nu}}^{\dagger}]_{11}} {\mathrm{Im}} \left\{ [{\bf
Y_{\nu}} {\cal M_{\nu}}^{\dagger} {\bf Y_{\nu}}^T]_{11} \right\}.
\label{eps1}
\eea
The value of the CP asymmetry depends on the details of the model,
however, we will show that there exists a {\it model independent}
upper bound on the CP asymmetry with several interesting physical
consequences.

To derive the upper bound on $|\epsilon_1|$, we will use
the parametrization of the Yukawa couplings introduced in 
\cite{Casas:2001sr}. There, it was proved that the most general
Yukawa coupling that satisfies eq.(\ref{seesaw}) is given by
\bea
\label{Ynu}
{\bf Y_\nu}=\frac{1}{\langle H_u^0\rangle} 
D_{\sqrt{M}} R D_{\sqrt{m}} U^+ ,
\eea
where, in an obvious notation, $D_{\sqrt{A}}\equiv +\sqrt{D_{A}}$, and
$R$ is a (complex) orthogonal matrix. Substituting in eq.(\ref{eps1}),
one gets
\bea
\epsilon_1& \simeq & -\frac{3}{8 \pi}  
\frac{M_1}{\langle H_u^0\rangle ^2}
\frac{ \sum_j m_j^2~{\mathrm{Im}} (R_{1j}^2)}{\sum_j m_j |R_{1j}|^2}.
\label{eps2}
\eea
Then, using the orthogonality condition $\sum_j R^2_{1j}=1$, it is
straight-forward to show that
\bea
|\epsilon_1| \lsim \frac{3}{8 \pi} 
\frac{M_1}{\langle H_u^0\rangle^2} (m_3-m_1).
\label{strictbound}
\eea
(Notice that the upper bound, like the CP asymmetry, 
goes to zero when the light neutrinos
are degenerate in mass.) 
A hierarchical spectrum of right-handed neutrinos strongly suggests a
spectrum of left-handed neutrinos also hierarchical, otherwise a big
conspiracy would be needed between $\bf {Y_{\nu}}$ and ${\cal M}$ to
produce a non-hierarchical spectrum. Therefore, we can assume $m_3 \gg
m_1$, hence
\bea
|\epsilon_1| \lsim\frac{3}{8 \pi} 
 \frac{M_1m_3}{\langle H_u^0\rangle^2}.
\label{bound}
\eea
Equations (\ref{bound}) and (\ref{strictbound}) are the main results
of this letter. 
They are valid in the seesaw model with hierarchical
right-handed neutrino masses, and arbitrary Yukawa matrices.
A formula similar to eq.(\ref{bound}) can be found in
\cite{Barbieri:2000ma}, who estimate $\epsilon$ to be our bound, but
say that unspecified cancellations can allow it to be larger. The
limit eq.(\ref{bound}) is present in \cite{Hamaguchi:2001gw}, who use it
to argue in favour of non-thermal right-handed sneutrino production.
Numerically similar bounds have been found in
specific models \cite{Buchmuller:1999zf,everyone}.

An immediate consequence of eq.(\ref{bound}) is that the CP
asymmetry, $\epsilon$, and the lightest right-handed neutrino mass,
$M_1$, are not completely independent parameters, contrary to what is
often assumed in the literature.  $\epsilon$ clearly depends on $M_1$,
but also on the unknown couplings of the $\nu_R$, so it would seem
reasonable to parametrize the predictions of leptogenesis with
$\epsilon$, $M_1$ and $\widetilde m_1$, regarding them as independent
parameters. However, the upper bound eq.(\ref{bound}) sets a
constraint on this parametrization: for instance, if $\epsilon>
10^{-6}$, the reference value chosen by Buchm\"uller and Pl\"umacher(BP), 
then $M_1 > 4 \times 10^{9}$ GeV.

This bound is not significantly weakened as the right-handed neutrinos
become less hierarchical. If $M_3>2M_2$ and $M_2>2M_1$, then
\bea
|\epsilon_1| 
\lsim \frac{1}{2 \pi} 
\frac{M_1}{\langle H_u^0\rangle^2}
 \frac{1}{[{\bf Y_{\nu} Y_{\nu}}^{\dagger}]_{11}} {\mathrm{Im}}
 \left\{  [{\bf Y_{\nu}} {\cal M}^{\dagger} {\bf Y_{\nu}}^T]_{11}
 \right\} 
\leq  \frac{1}{2 \pi} \frac{M_1}{\langle H_u^0\rangle^2}(m_3-m_1),
\label{eps1-degen}
\eea
i.e., a factor 4/3 larger than the bound eq.(\ref{strictbound}).  As
before, for hierarchical left-handed neutrinos, $m_3-m_1$ can be
safely approximated by $m_3$. Nonetheless, this mild hierarchy of
right-handed masses could be compatible with degenerate left-handed
neutrinos, with a certain amount of fine-tuning.  In that case,
$m_3-m_1 \simeq \Delta m^2_{atm} / 2 m_3$, hence, the maximum CP
asymmetry decreases as the overall scale of neutrino masses increases.

So far, we have not implemented in our bound the out-of-
equilibrium condition, 
\bea
\widetilde m_1= \sum_j m_j |R_{1j}|^2  < 5 \times 10^{-3} {\rm eV}.
\label{mtilde-R}
\eea
When one does this, the bound on $\epsilon$ only becomes slightly
stronger, and the improvement is numerically irrelevant.  Therefore,
and for the sake of clarity, we will use eq.(\ref{bound}) in the
forthcoming discussion.

Notice that from eq.(\ref{mtilde-R}) we can obtain an upper bound on
the lightest {\it left-handed} neutrino mass, $m_1$.  There is a
well-known bound on $\widetilde m_1$,
eq.(\ref{mtilde}), from requiring that lepton number violating
$\nu_{R_1}$ decays be out of equilibrium at $T \sim M_1$ (to avoid
washing-out any lepton asymmetry present at that time). The bound
$\widetilde m_1 < 5 \times 10^{-3}$ eV is usually applied to $m_1$,
assuming that $\widetilde m_1 \simeq m_1$.  This is not immediately
obvious; $m_1$ is the $\nu_{L_1}$ mass, and $\widetilde m_1$ the
rescaled $\nu_{R_1}$ decay rate. However, using the orthogonality of
$R$, it is clear that $m_1 \leq \widetilde m_1$, so this assumption is
justified.  This upper limit implies that, in the minimal seesaw model
considered here, leptogenesis cannot generate the baryon asymmetry if
the $\nu_L$ are degenerate\footnote{This statement supposes that the
Universe is radiation dominated when the $\nu_R$ decay. If it is
matter dominated ($e.g.$ by a scalar condensate), leptogenesis might
be possible.} \cite{Fischler:1991gn}.  More complicated models are
required \cite{Chun:2001dr}.

\section{Lower bound on the lightest right-handed neutrino mass}

In this section we will derive a bound on $M_1$ using the lower bound
on the CP asymmetry, and the information available on neutrino masses
and cosmology, for the case of hierarchical right-handed and
left-handed neutrinos. From eqs.(\ref{etaL},\ref{BAU},\ref{bound}),
one obtains
\bea
\label{boundM1}
M_1 \gsim \eta_B \frac{1-C}{C} 
\left[ \frac{n_{\nu_R}+ n_{\tilde{\nu}_R}}{s} ~ 
\frac{3}{8 \pi} \frac{m_3}{\langle H_u^0\rangle^2}~ \delta \right]^{-1}.
\eea
In this inequality, $\eta_B$ is constrained by Big Bang
Nucleosynthesis to lie in the range $(0.3-0.9) \times 10^{-10}$, and
$m_3 \simeq \sqrt{\Delta m^2_{atm}}$ by atmospheric neutrino data to
be within $0.04-0.08~{\rm eV}$.  The washout parameter, $\delta
\lappeq 1$, should be calculated case by case by integrating the full
Boltzmann equations \cite{Plumacher:1997kc}. On the other hand, the
remaining quantity, the right-handed neutrino density over the entropy
density, depends crucially on the mechanism of generation of
right-handed neutrinos: thermal production, or non-thermal production
(for instance during preheating) lead to different bounds on
$M_1$. Let us discuss each case separately.

\vspace{0.3cm}
\noindent
{\em Thermal production}

\vspace{0.2cm}

\noindent

In this case, the right-handed neutrinos are generated by scatterings
in the thermal bath. When the number density of right-handed neutrinos
is thermal at $T > M_1$, the prediction for the ratio of $n_{\nu_R}$
to the entropy density, $s$, is $n_{\nu_R}/s \sim 0.4/g_*$, where $g_*
\simeq 230$ is the number of propagating states in the supersymmetric
plasma.  Assuming that the asymmetry was not washed-out after being
generated ($\delta \sim 1$) gives a conservative lower bound of $M_1
\gsim 10^{8}$ GeV.  However, the numerical results of BP suggest that
this is improbable: if the Yukawa couplings are large enough to
produce a thermal density $n_{\nu_R}$, then the asymmetry will be
partially washed-out by lepton number violating interactions after the
decay of the right-handed neutrinos. Therefore, scaling our bound by $
n_{\nu_R}/s ~ \delta \sim 0.04/g_*$
\cite{Buchmuller:1999cu,Plumacher:1997kc}, we obtain
\beq
M_1 \gsim  10^{9} \left(\frac{\eta_B}{5 \times 10^{-11}} \right)
\left(\frac{.06 eV}{m_3} \right)
\left(\frac{2 \times 10^{-4}}{n_{\nu_R}/s ~ \delta}\right)
{\rm ~GeV} . 
\label{M1-thermal-conservative}
\eeq

The thermal production of right-handed neutrinos requires a 
reheat temperature larger than $M_1$. A typical value might
be $T_{reh} \sim 10~M_1$
\cite{Buchmuller:1999cu,Plumacher:1997kc}, so  we get 
\beq
T_{reh} \gsim 10^{10} 
\left(\frac{\eta_B}{5 \times 10^{-11}} \right)
\left(\frac{.06 eV}{m_3} \right)
\left(\frac{2 \times 10^{-4}}{n_{\nu_R}/s ~ \delta} \right)
\left( \frac{T_{reh}}{10 M_1} \right) {\rm ~GeV},
\label{lowTreh}
\eeq
or, using $T_{reh} > (1-10) M_1$,
\beq
T_{reh} \gsim 10^{8} -  10^{10} {\rm ~GeV~~~~   (baryon~ asymmetry)},
\label{lowbd}
\eeq
being more probable the large values in this range, since they take
into account the unavoidable washout effects. The bound
(\ref{lowTreh}) applies to hierarchical right-handed neutrinos,
irrespective of the form of the Yukawa matrix. It is the same as the
estimate made in \cite{Buchmuller:1999zf} for specific texture models.

It is enlightening to compare this lower bound on the reheat
temperature with the upper bound obtained from gravitino
overproduction \cite{gravitino,ENS}.  Gravitinos are produced by
scattering in the thermal bath at a rate~$\sim T^3/m_{pl}^2$, and then
decay with a lifetime~$\sim m_{pl}^2/m_{\tilde{G}}^3$.  Their decay
products can disassociate light elements, jeopardizing the successful
predictions of Big Bang Nucleosynthesis \cite{Lindley:1985bg}. To
prevent this, gravitinos should not be abundantly produced, and this
in turn imposes an upper bound on the reheat temperature
\cite{Kawasaki:1995af}:
\beq
T_{reh} \lsim  10^{9} - 10^{12} {\rm ~GeV~~~~   (gravitino~ production)}.
\label{grav26}
\eeq 
The first value, $T_{reh} \sim 10^{9}$ GeV, is the most commonly used
in the literature, although the weaker bound $T_{reh} \sim 10^{12}$
GeV, corresponding to 3 TeV $< m_{\tilde{G}} <$ 10 TeV, cannot be
precluded.  Gravitinos can also be produced efficiently in the
oscillations of the inflaton \cite{Kallosh:2000jj}, if the inflatino
mixes significantly with the gravitino \cite{Nilles:2001ry}.  This
leads to a stronger, although model-dependent, bound on $T_{reh}$,
which we do not consider.

Interestingly enough, there is a potential conflict between equations
(\ref{lowbd}) and (\ref{grav26}).  If the right-handed neutrinos are
produced thermally, the preferred gravitino bound $T_{reh} \lsim
10^{9}$ GeV is below the preferred leptogenesis bound $T_{reh} \gsim
10^{10}$ GeV.  There are various solutions to this possible
problem. If the gravitino is heavy, $T_{reh} > 10^{10}$ GeV is
consistent with BBN.  It is also possible to construct models where
the gravitino is the LSP, that allow reheat temperatures $< 10^{11}$
GeV \cite{ENS,Bolz:1998ek}. Alternatively, the number of right-handed
neutrinos produced thermally may be insufficient to generate the
baryon asymmetry.  In this case, other (non-thermal) production
mechanisms should come into play \cite{Hamaguchi:2001gw}.

\vspace{0.3cm}
\noindent
{\em Non-thermal production}

\vspace{0.2cm}

\noindent

Right-handed neutrinos could be coupled to the inflaton, and hence
produced perturbatively \cite{Asaka:1999yd} or non-perturbatively
\cite{Giudice:1999fb,Garcia-Bellido:2001cb} in inflaton decay. A
right-handed sneutrino condensate could also be generated by the
Affleck-Dine mechanism \cite{Hamaguchi:2001gw}.  In these cases, $M_1
<T_{reh}$ is not required.  Then, it is clear that the gravitino
problem will be easily avoided, since it is possible to create rather
heavy particles with relatively low reheat temperatures, that do not
endanger the predictions of Big Bang Nucleosynthesis.

There is nonetheless a bound on $M_1$ in this scenario, and on the
temperature of the Universe when the $\nu_R$ decay, $T_\Gamma< M_1$.
This has been considered in \cite{Hamaguchi:2001gw}, for the decay of
a $\tilde{\nu}_R$ condensate. We briefly review their discussion here,
applied to the general case of non-thermally produced ${\nu}_R$ or
$\tilde{\nu}_R$. We assume a distribution of right-handed neutrinos
that decay instantaneously, at a time equal to the right-handed
neutrino lifetime, to radiation that dominates the Universe.  The
decay products are thermalized because they have gauge
interactions. Probably, the right-handed neutrinos only contribute a
fraction of the energy density of the Universe, $\rho_U$, when they
decay; however, the baryon to entropy ratio will be maximum if the
right-handed neutrinos dominate the Universe. Hence, if we assume
$\rho_U \simeq \rho_{\nu_R}$, we will obtain a lower bound on
$\epsilon_1$ and therefore on $M_1$.  When the right-handed neutrinos
decay
\beq
 M_1~ n_{\nu_R} 
 \simeq  \frac{g_* \pi^2}{30}T_{\Gamma}^4 \sim s~ T_{\Gamma},
\eeq
which combined with eq.(\ref{boundM1}) gives   \cite{Hamaguchi:2001gw}
\bea
M_1>T_{\Gamma} \gsim 5 \times 10^{5} 
\left(\frac{\eta_B}{5 \times 10^{-11}} \right)
\left(\frac{.06 eV}{m_3} \right) ~~{\rm GeV}.
\label{Tdec1}
\eea
There is a lower bound on $T_{\Gamma}$, because if $M_1$ is large
(implying large $\epsilon_1$) but $T_{\Gamma}$ is small, then the
entropy produced per $\nu_R$ decay is large, and dilutes the
asymmetry.  The bound on $T_{\Gamma}$ sets the minimum temperature
possible in the Universe at the time that the lepton asymmetry is
generated. Consequently, it also represents a bound on the temperature
at which all the unwanted relics (gravitinos, moduli,...) cannot be
overproduced, since any relics produced after this moment cannot be
diluted by entropy production (otherwise, the baryon asymmetry would
also be diluted).

A particular example of non-thermal $\nu_R$ production is to create
the population of right-handed neutrinos during preheating. The
previous lower limit on $M_1$, eq.(\ref{Tdec1}), applies to this
scenario. In addition, the lower bound on $\epsilon_1$ from
eq.(\ref{bound}) implies that neutrinos must be relatively strongly
coupled to the inflaton in the model of \cite{Giudice:1999fb}, as we
will now see.

Reference \cite{Giudice:1999fb} assumes a model of chaotic inflation,
where the right-handed neutrino interacts with the inflaton, $\phi$,
via $g \phi \nu_R \nu_R$.  The effective mass of the $\nu_R$ is $M + g
\phi(t)$, so, as $\phi$ oscillates, it goes through zero for
sufficiently large $\phi$ oscillations. Significant numbers of $\nu_R$
can be produced while they are effectively massless.  The energy
density in right-handed neutrinos, divided by the energy density of
the inflaton, is of order \cite{Giudice:1999fb}
\beq
\frac{\rho_{\nu_R}}{\rho_{\phi}} \simeq \frac{4}{3 \pi^2} 
\frac{m_{\phi}^2}{\phi_0^2} q  =  \frac{g^2}{3 \pi^2},
\eeq
where $m_{\phi} \sim 10^{-6} m_{pl} $ is the oscillation frequency of
the inflaton, $\phi_0 \simeq m_{pl}/3$ is its initial value at the
start of oscillations, and $q \equiv g^2 \phi_0^2/m_{\phi}^2$.  The
final asymmetry can be estimated as \cite{Giudice:1999fb}
\beq
\frac{n_{B-L}}{s} \simeq 9 \times 10^{-8}~ \epsilon_1 ~ 
\frac{T_{reh}}{10^9 GeV}
 \frac{10^{15}GeV}{M_1} \frac{q}{10^{10}}.
\eeq
Implementing our bound in this estimate, we find
\beq
\frac{n_{B-L}}{s} \lappeq  2 \times  10^{-8} \frac{T_{reh}}{10^9 GeV}
  \frac{q}{10^{10}},
\eeq
so $q \gappeq 10^8$ is required. This corresponds to $g \gappeq .06$.

\vspace{0.3cm}
\noindent
{\em Loopholes}

\vspace{0.2cm}

\noindent

There are various ways of evading our lower bound on $M_1$ and
$T_{reh}$.

The formula we use for $\epsilon_1$ is applicable to hierarchical
right-handed neutrino masses: $M_3 \gg M_2 \gg M_1$.  The bound does
not apply for quasi-degenerate $\nu_R$ (see $e.g.$
\cite{Pilaftsis:1998pd}), where leptogenesis is complicated by mixing
among the $\nu_R$. However, the right-handed masses must be quite
close for the bound on $\epsilon_1$ to weaken: as shown in Section
\ref{sec2}, the upper bound on $\epsilon_1$ is only 4/3 larger when $M_3
> 2 M_2$, $M_2 > 2 M_1$.

In supersymmetric models, the baryon asymmetry can be generated via
the Affleck-Dine mechanism \cite{Affleck:1985fy}.  The magnitude of the
asymmetry in this scenario is controlled by a different combination of
parameters; we have not considered any possible implications our bound
might have.

Finally, we note that in non-supersymmetric models, there is no upper
bound on the reheat temperature from gravitino overproduction.  So,
the lower bound on $T_{reh}$ from leptogenesis is not disturbing.
However, the seesaw without SUSY suffers from a hierarchy problem.

\section{Summary}

In leptogenesis scenarios, the baryon asymmetry is generated in the
out-of-equilibrium decay of right-handed neutrinos and sneutrinos in
the early Universe.  The asymmetry is proportional to $\epsilon_1$,
which parametrizes CP violation in these decays.

We have shown that there exists a model independent
 upper bound on $\epsilon_1$ as a function
of $M_1$:
\bea
|\epsilon_1| \lsim\frac{3}{8 \pi} 
 \frac{M_1}{\langle H_u^0\rangle^2} (m_3- m_1) .
\label{bdc}
\eea
where $m_3$ is the heaviest light neutrino mass (we take 
$(m_3 - m_1 ) \simeq \sqrt{\Delta m^2_{atm}}$), 
and $M_1$ is the lightest right-handed
neutrino mass.  This bound assumes hierarchical right-handed neutrino
masses.

The observed baryon asymmetry $\eta_{B-L} \simeq 10^{-10}$ sets a
lower limit on $\epsilon_1$, and therefore on $M_1$.  If the
right-handed neutrinos are thermally produced, then $\eta_{B-L}
\lappeq 10^{- (2 - 3)} \epsilon_1$ and $M_1 > 10^{8-9}$ GeV. The reheat
temperature must be at least as large as $M_1$, so this implies
$T_{reh} > 10^{8} - 10^{10} $ GeV.  Numerical and analytic results
\cite{Buchmuller:1999cu,Plumacher:1997kc} suggest that 
$\eta_{B-L} \lappeq 10^{-3} \epsilon_1$ for $T_{reh} = 10 M_1$. These
parameters correspond to the lower bound $T_{reh} > 10^{ 10}$ GeV,
that in some scenarios conflicts with the upper bound from gravitino
overproduction.

We briefly discussed various loopholes in the limit on $\epsilon_1$, and
in the bound derived from it on $T_{reh}$.  If the right-handed
neutrinos are produced by a non-thermal process, $M_1$ can be much
larger than $T_{reh}$, so there is no lower bound on $T_{reh}$.  There
are analytic approximations \cite{Giudice:1999fb} for the lepton
asymmetry due to right-handed neutrinos generated during preheating;
implementing the bound (\ref{bdc}) 
in these formulae, we find that very efficient
$\nu_R$ production ($q > 10^8$) is required to get a large enough
asymmetry.

\section*{Acknowledgments}

We would like to thank Marco Peloso, Michael Pl\"umacher, Graham Ross
and Alessandro Strumia for comments and discussions.

\end{document}